\documentclass[letterpaper, 10 pt, conference]{ieeeconf} 
\IEEEoverridecommandlockouts
\usepackage{graphics} 
\usepackage{epsfig} 
\usepackage{mathptmx} 
\usepackage{times} 
\usepackage{amsmath} 
\usepackage{amssymb}  

\usepackage{subcaption}
\usepackage{xcolor}
\usepackage[cmintegrals]{newtxmath}

\begin{document}
\title{\LARGE \bf
Control Co-Design for Buoyancy-Controlled MHK Turbine: A Nested Optimization of Geometry and Spatial-Temporal Path Planning
}

\author{Arezoo~Hasankhani$^{1}$, Yufei~Tang$^{1}$, Austin Snyder$^{2}$, James~VanZwieten$^{3}$, Wei Qiao$^{4}$
\thanks{*This work was supported in part by the US National Science Foundation (NSF) through Grants ECCS-1809164 and CNS-1950400.}
\thanks{$^{1}$A. Hasankhani and Y. Tang are with the Department of Computer \& Electrical Engineering and Computer Science, Florida Atlantic University, Boca Raton, FL 33431, USA {\tt\small  \{ahasankhani2019, tangy\}@fau.edu}}
\thanks{$^{2}$A. Snyder is with the Department of Computer Science, University of North Carolina at Chapel Hill, Chapel Hill, NC 27514, USA {\tt\small adsnyder@live.unc.edu}}
\thanks{$^{3}$J. VanZwieten is with the Department of Civil, Environmental, and Geomatics Engineering, Florida Atlantic University, Boca Raton, FL 33431, USA {\tt\small jvanzwi@fau.edu}}
\thanks{$^{4}$W. Qiao is with the Department of Electrical and Computer Engineering, University of Nebraska-Lincoln, Lincoln, NE 68588, USA {\tt\small wqiao@engr.unl.edu}}
}

\maketitle
\thispagestyle{empty}
\pagestyle{empty}


\begin{abstract}
Recent research progress has confirmed that using advanced controls can result in massive increases in energy capture for marine hydrokinetic (MHK) energy systems, including ocean current turbines (OCTs) and wave energy converters (WECs); however, to realize maximum benefits, the controls, power-take-off system, and basic structure of the device must all be co-designed from early stages. This paper presents an OCT turbine control co-design framework, accounting for the plant geometry and spatial-temporal path planning to optimize the performance. Developing a control co-design framework means that it is now possible to evaluate the effects of changing plant geometry on a level playing field when accounting for the OCT plant power optimization. The investigated framework evaluates the key design parameters, including the sizes of the generator, rotor, and variable buoyancy tank in the OCT system, and formulates these parameters' effect on the OCT model and harnessed power through defining a power-to-weight ratio, subject to the design and operational constraints. The control co-design is formulated as a nested optimization problem, where the outer loop optimizes the plant geometry and the inner loop accounts for the spatial-temporal path planning to optimize the harnessed power with respect to the linear model of the OCT system and ocean current uncertainties. Compared with a baseline design, results verify the efficacy of the proposed framework in co-designing an optimal OCT system to gain the maximum power-to-weight ratio.
\end{abstract}


\section{INTRODUCTION}
An integrated design process is a crucial task when developing a multi-layer complex system such as marine hydrokinetic (MHK) energy turbines, which consist of hydrodynamic, mechanical, electrical, and cyber layers. The controller, which is a major part of the cyber layer, should be designed to complement the physical system design as an approach named control co-design \cite{garcia2019control}, thereby limiting adverse effects on the system efficiency, cost, and stability. The control co-design, integrated design process that has been followed is a so-called \emph{wave to wire} design in a sister field of wave energy conversion \cite{o2017co,coe2020initial,lyu2019optimization,herber2013wave,herber2014dynamic}. Due to the high investment and maintenance costs \cite{freeman2021rotor} of an \emph{ocean current turbine (OCT)}, a subset of MHK turbines specifically designed for open ocean applications, it is crucial to design an integrated OCT system, avoiding any misestimation in the subsystem design.

Developing a mechanically functional and economically feasible design for the OCT system has been recently gained a fair amount of attention. To maintain an operating depth of the OCT, this system has been designed based on various technologies, including variable buoyancy \cite{hasankhani2021modeling,coiro2017development,IHI2018development,wu2019dynamics}, lifting surfaces or wings \cite{Tandon,vanzwieten2006design,Minesto}, sub-sea winches \cite{baheri2018iterative}, and surface buoys \cite{shirasawa2016experimental}. Among these technologies, this paper will advance the buoyancy-controlled OCT presented in \cite{hasankhani2021modeling}, interpreted as a highly probable option to further with industrial applications \cite{IHI2018development}. In a general format, the mechanical and physical design has only entailed the prerequisites, including functionality, safety, reliability, size, and weight, neglecting the physical design's influence on the controllers. 

The controllers and cyber layers are primarily responsible for delivering optimized and safe power extracted from stochastic ocean currents by the OCT. To harness the maximum mechanical power, it is essential to deal with a path planning algorithm (as a part of the cyber layer) considering a spatial-temporal uncertain oceanic environment. The vast majority of the literature has observed path planning as an independent task from mechanical design. For a specific application of OCT, abundant advanced approaches have been proposed to cope with the path planning algorithm and dynamic ocean currents, such as model predictive control- \cite{bin2019centralized}, iterative learning- \cite{cobb2021iterative}, dynamic programming- \cite{reed2020optimal}, and reinforcement learning-based \cite{hasankhani_ccta} methods. However, the existing literature lacks an integrated design for power maximization subject to the plant geometry, dynamic response, and path control. 

Given the complexities of the OCT system and spatial-temporal uncertainties of ocean currents, a detailed study on designing the path controller is well deserved; however, the control challenges are intensified due to an inefficient mechanical design. Note that the planned path should qualify for the physical design, and the OCT system is able to follow this path through its controllers and actuators. Eventually, there exist several major features that are required to be addressed in the mechanical design to harness a fair amount of power from the ocean currents. Despite the intensive study on the OCT mechanical and structural design, a scientific gap in developing a co-design to couple the physical design and the cyber controller has been identified. There exist limited literature on the co-design of the MHK energy system, such as an ocean kite \cite{naik2021fused}; however, to the best of the authors' knowledge, it is the first time that this co-design problem is investigated for the OCT system.

In this paper, a turbine geometric/spatial-temporal control co-design is developed to maximize the power-to-weight ratio of a buoyancy-controlled OCT and analyze the whole design from ocean currents to power generation to ensure that it is optimal. To formulate a bi-directional coupling between plant design and path control, a nested co-optimization framework, consisting of two optimization loops is proposed: (i) an outer optimization loop responsible for the optimal design parameters addressed through a genetic algorithm (GA); and (ii) an inner optimization loop taking care of path planning solved by a model predictive control (MPC). The whole framework accounts for the dynamic model of the OCT system and spatial-temporal uncertainties that arise from the ocean current (see \cite{hasankhani2021modeling} for details), thereby the mechanical, electrical, and controller are optimized as an integrated unit (instead of designing each part individually) to maximize power harvesting and minimize the OCT weight. 

The remainder of this paper is organized as follows. Section~\ref{sec:prob. model} presents the OCT system model and its key design parameters. Section~\ref{sec:prop. method} describes the proposed control co-design architecture for the OCT system. Section~\ref{sec:results} shows the numerical results, and Section~\ref{sec:conclusion} presents the conclusion and future work.

\begin{figure}[t]
\centering
\includegraphics[width=0.92\linewidth]{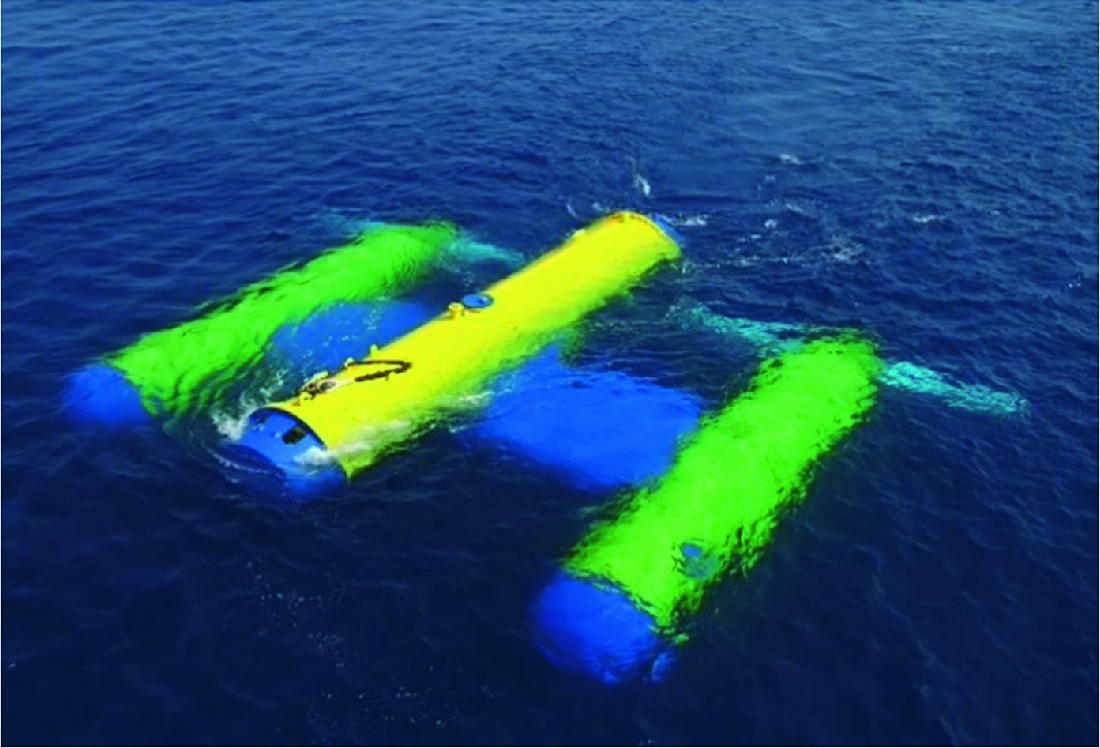}
\caption{Example of a buoyancy-controlled OCT that has counter-rotating rotor blades with a single variable buoyancy module. Image credit IHI Inc. \cite{ueno2018development,dodo2019development}.}
\label{fig:real_OCT}
\end{figure}

\section{PROBLEM DESCRIPTION} \label{sec:prob. model}
\subsection{OCT Background}
\begin{figure}[t]
\centering
\includegraphics[width=0.92\linewidth]{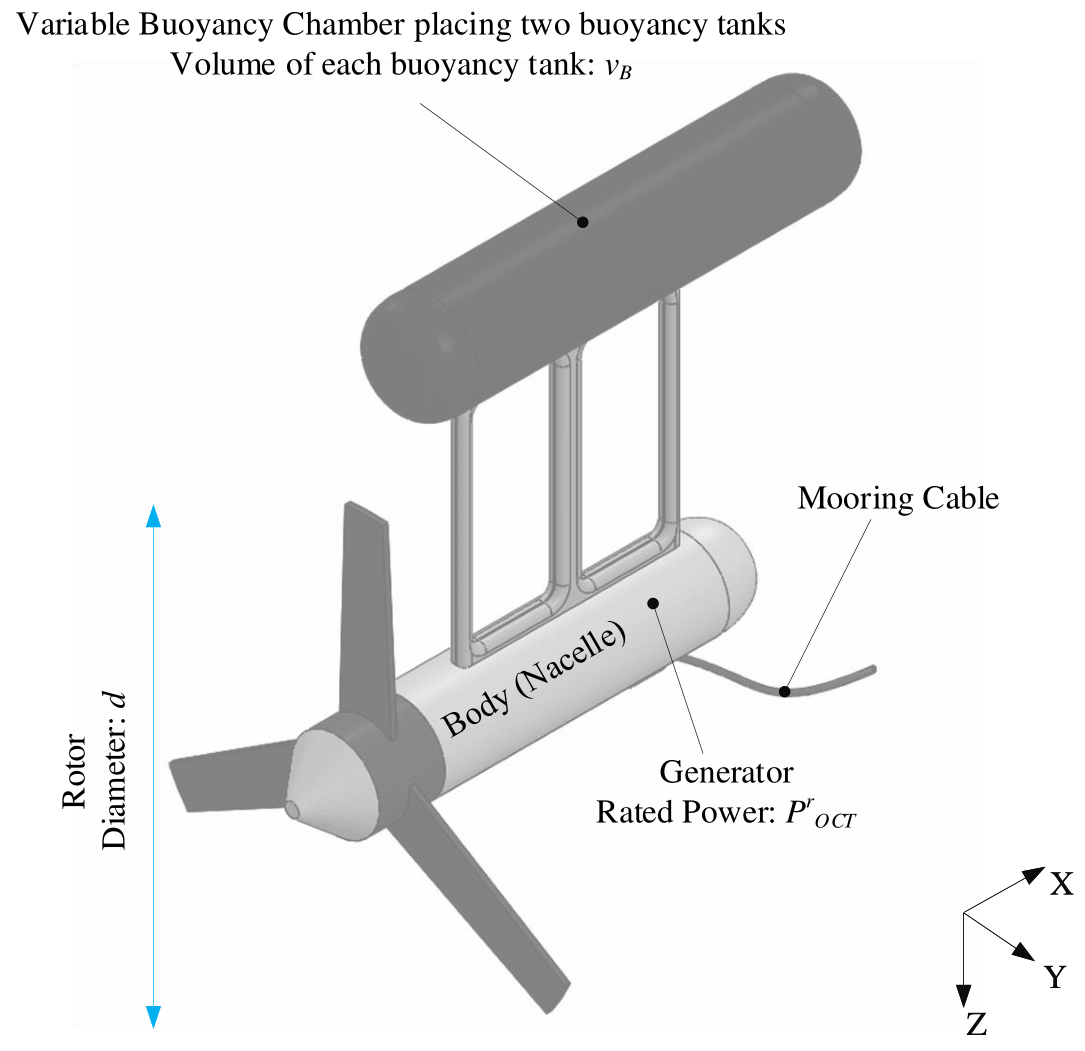}
\caption{Schematic diagram of the buoyancy-controlled OCT numerically simulated in this paper \cite{hasankhani2021modeling}.}
\label{fig:OCT}
\end{figure}

This work focuses on optimizing an OCT system that was initially designed with a rated power of 700 $\mathrm{kW}$ for harnessing power from the Gulf Stream off Florida's East Coast following an OCT prototype from IHI Corp. \cite{ueno2018development}, shown in Fig. \ref{fig:real_OCT}. The turbine was designed such that it will operate at a depth of 50 $\mathrm{m}$ in a homogeneous current speed of 1.6 $\mathrm{m/s}$ assuming 50\% filled buoyancy tanks. A schematic of the investigated OCT is shown in Fig. \ref{fig:OCT}, consisting of a body tethered to an anchor via a 607 $\mathrm{m}$ mooring cable and a variable buoyancy tank to control its vertical motion \cite{hasankhani2021modeling}.

\subsection{OCT Dynamic Modeling for Control} \label{sec:OCT model}
To formulate the OCT's kinematics, this system is described through five coordinate frames, including the body-fixed $(\mathcal{T}_B)$, the inertial $(\mathcal{T}_I)$, the momentum mesh $(\mathcal{T}_M)$, the shaft $(\mathcal{T}_S)$, and multiple rotor blade $(\mathcal{T}_R)$ frames.


 
\textbf{Equations of Motion:} Seven degree-of-freedom (DOF) equations of motion are created to describe the dynamic model of the OCT system as suggested in \cite{vanzwieten2012numerical}, with all forces and moments are shifted to the $\mathcal{T}_{\mathrm{B}}$ frame. The system is represented by 14 states, including the position of OCT body $\mathcal{P}=[x~y~z]$, linear velocity of the OCT body $\mathcal{\dot{P}}=[u~v~w]$, Euler angles of the OCT body $\Theta=[\phi~\theta~\psi]$, angular velocity of the OCT body $\dot{\Theta}=[p_{\mathrm{b}}~q~r]$, angular velocity of the rotor $p_{\mathrm{r}}$, and rotation angle of the rotor blade $\phi_{\mathrm{r}}$. These equations couple a 6-DOF motion of the OCT body,
\begin{align}\label{EOM}
    \begin{bmatrix}
            \mathcal{\ddot{P}}     \\
            \ddot{\Theta}
    \end{bmatrix}
    =\mathcal{M}^{-1} \mathcal{F}
\end{align}
where
\begin{equation}\label{equation of motion_1}
\resizebox{0.43\textwidth}{!}{$
    \mathcal{M}= \begin{bmatrix}
    m & 0 & 0 & 0 & m_{\mathrm{b}} z_{\mathrm{cg_{b}}} & 0 \\[2ex]
    0 & m & 0 & -m_{\mathrm{b}} z_{\mathrm{cg_{b}}} & 0 & m x_{\mathrm{cg}} \\[2ex]
    0 & 0 & m & 0 & -m x_{\mathrm{cg}} & 0 \\[2ex]
    0 & -m_{\mathrm{b}} z_{\mathrm{cg_{b}}} & 0 & I_{x_\mathrm{{b}}} & 0 & -I_{x z_{\mathrm{b}}} \\[2ex]
    m_{\mathrm{{b}}} z_{\mathrm{cg_{b}}} & 0 & -m x_{\mathrm{cg}} & 0 & I_{y} & 0 \\[2ex]
    0 & m x_{\mathrm{{cg}}} & 0 & -I_{x z_{\mathrm{{b}}}} & 0 & I_{z}
\end{bmatrix}$}
\end{equation}
\begin{equation}\label{equation of motion_2}
\resizebox{0.43\textwidth}{!}{$
    \mathcal{F}= \begin{bmatrix}
    f_{x}+m(v r-w q)+m x_{\mathrm{cg}}\left(q^{2}+r^{2}\right)-m_{\mathrm{b}} z_{cg_{b}} p_{b} r \\[2ex]
    f_{y}-m u r+w\left(m_{\mathrm{b}} p_{\mathrm{b}}+m_{\mathrm{r}} p_{\mathrm{r}}\right) -m_{\mathrm{b}} z_{\mathrm{cg_{b}}} q r \\
    - m_{\mathrm{b}} x_{\mathrm{cg_{b}}} q p_{\mathrm{b}}-m_{\mathrm{r}} x_{\mathrm{cg_{r}}} q p_{\mathrm{r}} \\[2ex]
    f_{z}+m u q-v\left(m_{\mathrm{b}} p_{\mathrm{b}}+m_{\mathrm{r}} p_{\mathrm{r}}\right)+m_{\mathrm{b}} z_{\mathrm{cg_{b}}}\left(p_{\mathrm{b}}^{2}+q^{2}\right) \\
    -m_{\mathrm{b}} x_{\mathrm{cg_{b}}} r p_{\mathrm{b}}-m_{\mathrm{r}} x_{\mathrm{cg_{r}}} r p_{\mathrm{r}} \\[2ex]
    M_{x_{\mathrm{b}}}+\tau_{\mathrm{em}}-q r\left(I_{z_{\mathrm{b}}}-I_{y_{\mathrm{b}}}\right)+I_{x z_{\mathrm{b}}} p_{\mathrm{b}} q \\
    -m_{\mathrm{b}} z_{\mathrm{cg_{b}}}\left(w p_{\mathrm{b}}-u r\right) \\[2ex]
    M_{y}-r p_{\mathrm{b}}\left(I_{x_{\mathrm{b}}}-I_{z_{\mathrm{b}}}\right)-r p_{\mathrm{r}}\left(I_{x_{\mathrm{r}}}-I_{z_{\mathrm{r}}}\right)-I_{x z_{\mathrm{b}}}\left(p_{\mathrm{b}}^{2}-r^{2}\right) \\
    +m_{\mathrm{b}} z_{\mathrm{cg_{b}}}(v r-w q)-m x_{\mathrm{cg}} u q+m_{\mathrm{b}} x_{\mathrm{cg_{b}}} v p_{\mathrm{b}}+m_{\mathrm{r}} x_{\mathrm{cg_{r}}} v p_{\mathrm{r}} \\[2ex]
    M_{z}-q p_{\mathrm{b}}\left(I_{y_{\mathrm{b}}}-I_{x_{\mathrm{b}}}\right)-q p_{\mathrm{r}}\left(I_{y_{\mathrm{r}}}-I_{x_{\mathrm{r}}}\right)-I_{x z_{\mathrm{b}}} r q\\
    -m x_{\mathrm{cg}} \text { u } r+m_{\mathrm{b}} x_{\mathrm{cg_{b}}} w p_{\mathrm{b}}+m_{\mathrm{r}}^{v} x_{\mathrm{cg_{r}}} w p_{\mathrm{r}}
\end{bmatrix}$}
\end{equation}
and a 1-DOF rotation of the rotor about the x-axis of the body frame:
\begin{equation}\label{equation of motion_3}
    \dot{p}_{\mathrm{r}}=\frac{M_{x_{\mathrm{r}}}-\tau_{\mathrm{em}}-qr(I_{z_{\mathrm{r}}}-I_{y_{\mathrm{r}}})}{I_{x_\mathrm{r}}}
\end{equation}
where, $f_{(.)}$ denotes the force, and $M_{(.)}$ is the moment. Note that the mass $m_{(.)}$, the moment of inertia $I_{(.)}$, and the center of gravity ${(.)}_{\mathrm{cg}}$ include both the actual inertial properties and added inertial properties of the OCT (denoted as \emph{virtual}) as suggested in \cite{vanzwieten2012numerical}. $(.)_{x}$, $(.)_{y}$, and $(.)_{z}$ denote the portion $(.)$ about $x-$, $y-$, and $z-$ axes. $\tau_{\mathrm{em}}$ denotes the electromechanical shaft torque. Eventually, $(.)_{\mathrm{r}}$ and $(.)_{\mathrm{b}}$ refer to the rotor and the body.

The total forcing on the OCT system is modeled through the gravitational and buoyancy force $f_{\mathrm{gb}}$, rotor force $f_\mathrm{r}$, body force $f_\mathrm{b}$, and cable force $f_\mathrm{c}$, namely $f=f_{\mathrm{gb}}+f_\mathrm{r}+f_\mathrm{b}+f_\mathrm{c}$; the total moments are accordingly calculated as fully presented in \cite{vanzwieten2012numerical}. To avoid complexities arise when applying the nonlinear dynamic model, the OCT model is linearized for controller development, as detailed in the followings.

\textbf{Linear Model:}
To form the linear state space model for the OCT system, the equations of motion are averaged around the nominal condition (equilibrium point), thereby formulating the linear model as follows:
\begin{align}\label{linear model}
	\delta \dot{x}_{n\times 1} &= A_{n\times n} \delta x_{n\times 1} + B_{n\times m} \delta u_{m \times 1}
\end{align}

Here, the states $\delta x$ and control inputs $\delta u$ are defined as deviation from the nominal condition. $\delta x \in \mathbb{R}^{n}$ and $\delta u \in \mathbb{R}^{m}$ with $n$ and $m$ denoting the number of states and control inputs used by the linear model ($n=13$ and $m=3$ in our OCT system), namely:
\begin{equation}
    \delta x=[\delta u~\delta v~\delta w~ \delta p_b~ \delta p_r~ \delta q~ \delta r~ \delta x~ \delta y~ \delta z~ \delta \phi~ \delta \theta~ \delta \psi]
\end{equation}
\begin{equation}
    \delta u=[\delta B_\mathrm{f}~ \delta B_\mathrm{a}~ \delta \tau_{\mathrm{em}}]   
\end{equation}
where the control inputs are defined by forward buoyancy tank fill fraction $B_\mathrm{f}$, aft buoyancy tank fill fraction $B_\mathrm{a}$, and $\tau_{\mathrm{em}}$. The fill fractions are limited by the ratio between the buoyancy tank size $\nu_{\mathrm{B}}$ and the base buoyancy tank size $\nu_{\mathrm{B}}^{\mathrm{b}}$ (as formulated in (\ref{constraint2})). It is noted that the linear model has one less state than the non-linear model as the rotor rotation angle state, $\phi_{\mathrm{r}}$, is eliminated during the linearization process as described in \cite{ngo2020constrained}. Given that the nominal condition is defined by an averaged homogeneous flow speed of $v_{\mathrm{eq}}=1.6 \mathrm{m/s}$, the nominal states and control inputs are characterized by $x_{\mathrm{eq}}=[0~ 0~ 0~ 0~ 1.49~ 0~ 0~ 554.50~ 0.38~ 50~ 0.01~ 0.00~ 3.14]$ and $u_{\mathrm{eq}}=[0.4677~  0.4677~ -188280]$.

Note that among all the states the major focus here is on the vertical position (depth) $z$, e.g., vertical path planning, which is primarily controlled through the fill fractions $B_{(.)}$. To further reduce the complexity of the model, this work leverages a linear equation relating $z$ to $B_{(.)}$ based on the authors' previous work \cite{Hasankhani_spatiotemporal}. To control the operating depth, the OCT system should be able to change and hold the optimal depth. Hence, the fill fractions must be set such that the OCT system is placed in an optimal depth and maintains that depth when the flow velocities change. The fill fraction at time instant $k$ is calculated by:
\begin{equation}\label{Fill Fraction}
    B_{\mathrm{(.)}}(k+1) = B_{\mathrm{(.)}}(k) + \Delta B_{\mathrm{(.)}}^{\mathrm{HD}} + \Delta B_{\mathrm{(.)}}^{\mathrm{CD}}
\end{equation}
where $\Delta B_{\mathrm{(.)}}^{\mathrm{HD}}$ and $\Delta B_{\mathrm{(.)}}^{\mathrm{CD}}$ are the changes in the fill fractions to hold the operating depth and change the depth, respectively. These fill fraction changes are calculated assuming the linear equations between fill fraction changes and depth changes $\frac{dF_{\text{F}}}{dz}$ as well as flow speed changes and depth changes $\frac{dv}{dz}$, namely:  
\begin{equation}\label{Fill Fraction_HD}
    \Delta B_{\mathrm{(.)}}^{\mathrm{HD}}= \frac{dF_{\text{F}}}{dz}\frac{dz}{dv} \Delta v=\kappa_{1} \Delta v
\end{equation}
\begin{equation}\label{Fill Fraction_CD}
    \Delta B_{\mathrm{(.)}}^{\mathrm{CD}}= \frac{dF_{\text{F}}}{dz} \Delta z= \kappa_{2} \Delta z
\end{equation}
where $\kappa_{1}=\frac{dF_{\text{F}}}{dz}\frac{dz}{dv}$ and $\kappa_{2}=\frac{dF_{\text{F}}}{dz}$ are the constant coefficients, which are presented in Table \ref{table:dimension OCT}. Each fill fraction, $B_{(.)}$, is limited between 0 (chamber filled with water) and 1 (chamber filled with air).


\textbf{OCT Output Power Model:}
The net harnessed power from the OCT system has been previously formulated in \cite{Hasankhani_spatiotemporal}, consisting of three terms of (i) power generation $P_{\mathrm{OCT}}$; (ii) power consumption to hold the operating depth $P_{\mathrm{HD}}$; and (iii) power consumption to change the operating depth and relocate the OCT $P_{\mathrm{CD}}$, namely:
\begin{equation}\label{Power}
    P_{\mathrm{net}} = P_{\mathrm{OCT}} - P_{\mathrm{HD}} - P_{\mathrm{CD}}
\end{equation}
where
\begin{equation}\label{Power_OCT}
    P_{\mathrm{OCT}} = \min(\frac{1}{2}\rho \pi (\frac{d}{2})^2 v^{3} c_{\mathrm{p}}, P^{\mathrm{r}}_{\mathrm{g}})
\end{equation}
\begin{equation}\label{Pump Power HD}
    P_{\mathrm{HD}} =
    \begin{cases}
    0 ,&  \text{$\Delta v < 0$}\\
    \frac{\zeta(\kappa_{1} \Delta v) }{\Delta t},&  \text{$\Delta v > 0$}\\
\end{cases}
\end{equation}
\begin{equation}\label{Pump Power CD}
    P_{\mathrm{CD}} =
    \begin{cases}
    0 ,&  \text{$\Delta z > 0$}\\
    \frac{\zeta(\kappa_{2} \Delta z) }{\Delta t},&  \text{$\Delta z < 0$}\\
\end{cases}
\end{equation}
where $\rho$ is the water density, $d$ is the rotor diameter, $v$ is the ocean velocity, $c_{\mathrm{p}}$ denotes the average power coefﬁcient, $P^{\mathrm{r}}_{\mathrm{g}}$ is the rated power of the OCT, $\Delta t$ is the sampling time, and $\zeta$ denotes the constant coefficient. 

\subsection{OCT Plant Parameters for Design} \label{sec:design_parameters}
A prominent design of the OCT system may entail several prerequisites: (i) harnessing the maximum power while tethered to the anchor; (ii) conducting a real-time search to find the vertical position with the maximum power subject to consuming the minimal power to relocate the OCT system; and (iii) acquiring the maximum power with a minimum OCT weight. To address these conditions, the dominant design parameters of the OCT system should be pointed out:

\textbf{Rotor:} The rotor is a key parameter in the OCT system design, directly affecting the whole system design and power generation. To highlight the importance of the rotor, a complete study on the hydrodynamic design of the OCT's rotor (airfoil size and shape with the corresponding power and thrust coefficients) has been conducted in \cite{vanzwieten2016assessment}. On account of the results obtained through \cite{vanzwieten2016assessment}, the main focus in the current study is on \emph{rotor diameter}. Larger rotor diameters increase the power production from OCT systems operating in a homogeneous flow field in the order of the square of the rotor diameter (see (\ref{Power_OCT})), but with increased hydrodynamic system drag/thrust (also a function of the square of the diameter) and increased weight. The increased drag/thrust force on the rotor requires increased buoyancy forces to maintain the same elevation, which can be achieved by maintaining less water in the buoyancy tanks. However, for higher flow speeds, this increases the minimum achievable operating depth (i.e., depth when buoyancy tanks are completely filled with air \cite{hasankhani2021modeling}), resulting in a decrease in the available energy density as available energy is typically strongest near the sea surface.


\begin{figure}[t]
\centering
\includegraphics[width=0.98\linewidth]{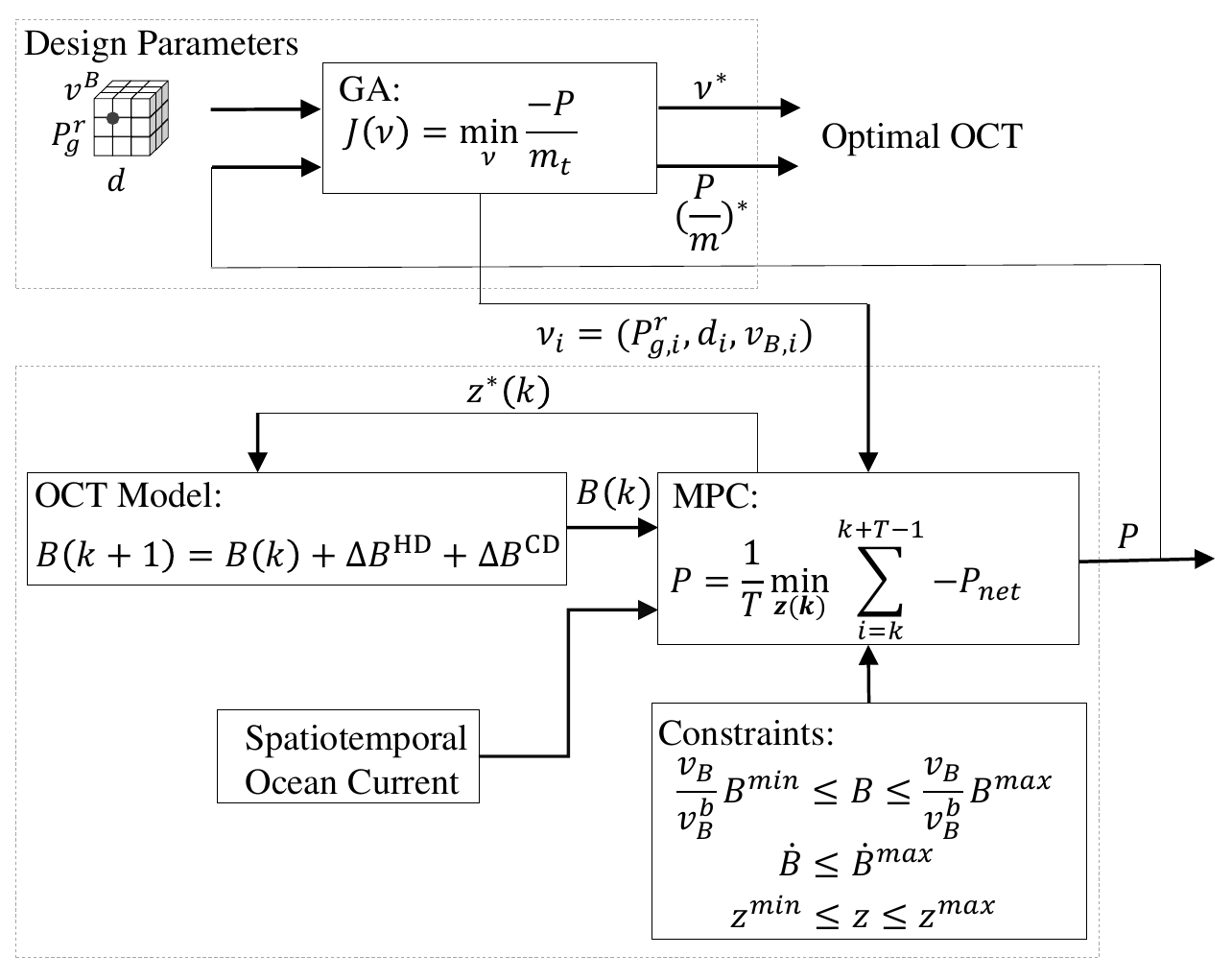}
\caption{Proposed control co-design framework for optimizing the power-to-weight ratio, while accounting for the OCT geometry and spatial-temporal path control.}
\label{fig:OCT co-design}
\end{figure}

\textbf{Buoyancy Tank:} In our investigated OCT system, the buoyancy tank is another important design parameter that keeps the system afloat. Accordingly, the operating depth of the tethered OCT plant is initially determined through the amount of water in the tank, thereby imposing constraints on the vertical forcing and movement based on tank size. The results obtained in the authors' previous work \cite{hasankhani2021modeling} suggest that the buoyancy tank size was not optimally designed since the OCT system can operate in a wide range of depth (e.g., $z=27$ $\mathrm{m}$ to 216 $\mathrm{m}$ for a flow speed of 2.5 $\mathrm{m/s}$) and can hit the depth of 50 $\mathrm{m}$ even with a strong flow speed of 2.5 $\mathrm{m/s}$. Note that in the buoyancy tank, a pump drives water while setting the tank's pressure at vacuum pressure ($\cong0$ $\mathrm{kPa}$). Therefore, the key design parameter is the \emph{tank size} (denoted as $\nu_{\mathrm{B}}$ for the volume of each of the two buoyancy tanks). A larger tank size allows for the larger movement of the OCT, but increases the system's weight and cost. 


\textbf{Generator:} To design a reliable and efficient OCT plant, the generator configuration plays an important role, limiting the power produced from the ocean currents (again, see (\ref{Power_OCT})). In the current control co-design framework, the primary concentration is on the \emph{generator rated power}. Picking a generator with a large rated power allows an electrical power output that is directly related to available hydrodynamic power from the ocean currents, however, at the expense of increased generator size, weight, and cost. Therefore, optimizing generator power rating to maximize the power-to-weight ratio (a surrogate for the cost of electricity) requires site specific energy resource data (commonly measured through the ``power density" \cite{Machado}).


\section{PROPOSED CONTROL CO-DESIGN FRAMEWORK FOR OCT} \label{sec:prop. method}
A nested control co-design framework is developed with the objective of maximizing the power-to-weight ratio of the OCT plant. The investigated architecture undertakes a combined geometric, structural design, and path planning to advance the performance of the OCT. Operating in a nested format, the framework consists of two main loops: the power-to-weight maximization outer loop addressed through the GA algorithm and the power-harnessing maximization inner loop solved by the MPC approach. A schematic of the proposed framework is shown in Fig. \ref{fig:OCT co-design}. The GA method is applied to vary the design parameters and find the optimal value for these design parameters. To start the GA, a set of initial design parameters (denoted as a population in the GA method) are randomly generated, which is passed to the inner loop of spatial-temporal power optimization to calculate its corresponding optimal power. To find the optimal OCT design, each population should be ranked according to a so-called \emph{fitness}, which is defined by the power-to-weight ratio. The MPC-based inner loop seeks to find a sequence of the optimal depths over prediction horizon $T$ to maximize the harnessed power from the OCT for each population received from the outer loop.

Specifically, the control co-design is formulated as an optimization problem that seeks to maximize the power-to-weight ratio (transferring from maximization to minimization through multiplying the objective function by $-1$), subject to the operational and design constraints:
\begin{equation}\label{MPC}
\begin{split}
        J(\upsilon) =  \underset{\upsilon}{\min}\frac{-P}{m_{\mathrm{t}}} 
\end{split}
\end{equation}
s.t.
\begin{equation}\label{constraint1}
    \upsilon^{\min }\leq \upsilon \leq \upsilon^{\max }
\end{equation}
\begin{equation}\label{MPC}
\begin{split}
        P =  \frac{1}{T}\underset{\mathbf{z}(k)}{\min}\sum_{i=k}^{k+T-1}[-P_{\mathrm{net}}(\upsilon,z(i|k))] 
\end{split}
\end{equation}
\begin{equation}\label{constraint6}
    m_{\mathrm{t}}=m^{\mathrm{b}}_{\mathrm{t}}+\Delta m_{\mathrm{r}}+\Delta m_{\mathrm{g}}+\Delta m_{\mathrm{B}}
\end{equation}

Assuming that the inner spatial-temporal power optimization (through vertical path planning) is formulated by:
\begin{equation}\label{MPC}
\begin{split}
        P =  \frac{1}{T}\underset{\mathbf{z}(k)}{\min}\sum_{i=k}^{k+T-1}[-P_{\mathrm{net}}(\upsilon,z(i|k))] 
\end{split}
\end{equation}
s.t.
\begin{equation}\label{constraint1}
    B_{\mathrm{(.)}}(k+1) = B_{\mathrm{(.)}}(k) + \Delta B_{\mathrm{(.)}}^{\mathrm{HD}} + \Delta B_{\mathrm{(.)}}^{\mathrm{CD}}
\end{equation}
\begin{equation}\label{constraint2}
    \frac{\nu_\mathrm{B}}{\nu^\mathrm{b}_\mathrm{B}}B_{(.)}^{\min}\leq B_{(.)}\leq \frac{\nu_\mathrm{B}}{\nu^\mathrm{b}_\mathrm{B}}B_{(.)}^{\max}
\end{equation}
\begin{equation}\label{constraint3}
    \dot{B}_{(.)}\leq \dot{B}_{(.)}^{\max}
\end{equation}
\begin{equation}\label{constraint4}
    z^{\min }\leq z(i|k) \leq z^{\max }
\end{equation}
where $\upsilon=[P^\mathrm{r}_{\mathrm{g}},~d,~\nu_{\mathrm{B}}]$ are decision variables as previously defined generator rated power, rotor diameter, and buoyancy tank size, respectively. Define $\mathbf{z}(k) \triangleq [z(k|k),...,z(k|k+T-1)]^{\text{T}}$. $m_{\mathrm{t}}$ is the total mass, consisting of the base mass, $m^{\mathrm{b}}_{\mathrm{t}}$, and mass deviations arising from changing the design parameters (i.e., rotor $\Delta m_{\mathrm{r}}$, generator $\Delta m_{\mathrm{g}}$, and variable buoyancy $\Delta m_{\mathrm{B}}$). Notation $z(i|k)$ represents the value of $z$ at time $i$ given its value at time $k$. Constraints (\ref{constraint2}) and ({\ref{constraint3}}) represent the limitations on the fill fractions. The major control objective of the spatial-temporal power optimization is to plan the vertical path for the OCT system, where constraint (\ref{constraint4}) limits the vertical movement of the system. Inspired by \cite{bafandeh2018comparative,nash2021robust}, the MPC-based spatial-temporal objective function in (\ref{MPC}) can be minimized using dynamic programming (DP) by forwarding recursion to find the global minimum over a predefined depth-time grid.

Here, the main challenge is to relate the mass deviation terms to the design parameters. To do this, two expressions initially formulated for the wind turbine's rotor and generator (recognized as one of the most similar plant to the OCT system) are extracted from the previous study \cite{fingersh2006wind}. To cope with the OCT system, it is important to set the coefficient $\alpha_1$ in the rotor mass expression (\ref{mass_r}) according to the increased mass of the OCT rotor in comparison with the wind turbine's rotor. In this study, the OCT's rotor is designed based on an FX-83-W hydrofoil (see \cite{vanzwieten2016assessment} for details), outperforming other rotor designs in generating more power $P_{\mathrm{OCT}}$. Hence, the mass deviations of the rotor and generator are:
\begin{equation}\label{mass_r}
    \Delta m_{\mathrm{r}} = \alpha_{1} (\frac{d_\mathrm{r}^\mathrm{b}}{2}+\frac{\Delta d}{2})^{\alpha_2} - m_{\mathrm{r}}^{\mathrm{b}}
\end{equation}
\begin{equation}\label{mass_g}
    \Delta m_{\mathrm{g}} = \beta_1 (p^{\mathrm{r,b}}_{\mathrm{g}}+\Delta p^{\mathrm{r}}_{\mathrm{g}})^{\beta_2} - m_{\mathrm{g}}^{\mathrm{b}}
\end{equation}

The equation for the variable buoyancy tank is formulated according to our previous numerical simulations study \cite{hasankhani2021modeling}, namely:
\begin{equation}\label{mass_vB}
    \Delta m_{\mathrm{B}} = \gamma_{1} \Delta \nu_{\mathrm{B}}
\end{equation}
where $\alpha_1$, $\alpha_2$, $\beta_1$, $\beta_2$, and $\gamma_1$ are the coefficients as presented in Table \ref{table:dimension OCT}.

\section{SIMULATION RESULTS and DISCUSSION} \label{sec:results}
\subsection{Simulation Setup}
The efficacy of the proposed control co-design framework is tested on a sample buoyancy-controlled OCT detailed in Section \ref{sec:OCT model}. The base design parameters suggested in \cite{hasankhani2021modeling} are presented in Table \ref{table:dimension OCT}. The simulations are fully conducted in Python on a machine equipped with a 2.3 $\mathrm{\,GHz}$ CPU and 32 $\mathrm{\,GB}$ of RAM. For the MPC-based spatial-temporal power optimization algorithm, the time step is set as 1 hour, the prediction horizon is set as $T=2$ hours, and the number of discrete depths is 17 within the allowable depth range (i.e., between $z_{\mathrm{min}}$ and $z_{\mathrm{max}}$ as presented in Table \ref{table:dimension OCT}). The design parameters are assumed to change within $\upsilon^{\min}=0.1 \times \upsilon^{b}$ and $\upsilon^{\max}=1.1 \times \upsilon^{b}$ to help ensure that the compatibility is maintained with the OCT components that are satisfied at their baseline values. Also, to introduce spatial-temporal uncertainties of the ocean currents into our model, we use a set of real measured ocean data at a latitude of $26.09 ^{\circ}N$ and longitude of $-79.80 ^{\circ}E$ by a 75 kHz acoustic Doppler current profiler (ADCP) with a vertical resolution of 5 $\mathrm{m}$ within 400 $\mathrm{m}$ \cite{Machado}.

\begin{table}[t]
    \caption{Parameters of the buoyancy-controlled OCT.}
    \centering
    \resizebox{\columnwidth}{!}{%
    \begin{tabular}{|c|c|c|c|}
        \hline
         Symbol & Description & Unit & Value \\
         \hline
        $d_\mathrm{r}^{\mathrm{b}}$ & Base rotor diameter & $\mathrm{m}$ & 20 \\
        $P^{\mathrm{r,b}}_{\mathrm{g}}$ & Base generator rated power & $\mathrm{kW}$ & 700\\        
        $\nu^{\mathrm{b}}_{\mathrm{B}}$ & Base volume of each buoyancy tank & $\mathrm{m^{3}}$ & 31.215\\
        $z^{\min }$ & Minimum vertical position & $\mathrm{m}$ & 50\\
        $z^{\max}$ & Maximum vertical position & $\mathrm{m}$ & 150\\
        $B^{\min }_{(.)}$ & Minimum buoyancy tank fill fraction & $-$ & 0\\
        $B^{\max }_{(.)}$ & Maximum buoyancy tank fill fraction & $-$ & 1\\
        $\dot{B}_{(.)}^{\max}$ & Maximum slew rate of fill fraction & $\mathrm{1/s}$ & $7.45 \times 10^{-4}$\\
        $m_{\mathrm{t}}^{\mathrm{b}}$ & Total mass & $\mathrm{kg}$ & 497800\\
        $m_{\mathrm{r}}^{\mathrm{b}}$ & Mass of base rotor & $\mathrm{kg}$ & 61573\\
        $m_{\mathrm{g}}^{\mathrm{b}}$ & Mass of base generator & $\mathrm{kg}$ & 2246.9\\
        $m_{\mathrm{B}}^{\mathrm{b}}$ & Mass of base buoyancy tank & $\mathrm{kg}$ & 20427\\
        $\zeta$ & Coefficient for power equation & $\mathrm{kWh}$ & 14.02\\
        $\kappa_{1}$ & Coefficient for power and fill fraction equations & $\mathrm{s/m}$ & 0.65\\
        $\kappa_{2}$ & Coefficient for power and fill fraction equations & $\mathrm{1/m}$ & -0.0026\\
        $\alpha_{1}$ & Coefficient for rotor mass equation & $\mathrm{kg/m}$ & 74.2832\\
        $\alpha_{2}$ & Coefficient for rotor mass equation & $-$ & 2.9158\\
        $\beta_{1}$ & Coefficient for generator mass equation & $\mathrm{kg/kW}$ & 5.34\\
        $\beta_{2}$ & Coefficient for generator mass equation & $-$ & 0.9223\\
        $\gamma_{1}$ & Coefficient for variable buoyancy mass equation & $\mathrm{kg/m^{3}}$ & 650.0721\\
        \hline
    \end{tabular}}
    \label{table:dimension OCT}
\end{table}

The following three cases are simulated and compared:
\begin{itemize}
    \item \textbf{Baseline OCT design:} The baseline OCT design investigated in this study is discussed in Section \ref{sec:prob. model}, and its major design parameters are presented in Table \ref{table:dimension OCT}. The mechanical and structural design of this system was detailed in \cite{hasankhani2021modeling}, which lacks an analysis to account for the structural design and its effect on the harnessed power. Here, the spatial-temporal optimization is performed for the baseline OCT system, and the maximum harnessed power along with the power-to-weight ratio for the baseline are found.
    \item \textbf{Co-design for a single OCT design parameter:} The co-design framework solves the optimization problem of maximizing the power-to-weight ratio for a single OCT design parameter. Three separate optimizations are performed for the buoyancy tank, generator, and rotor as major design parameters.
    \item \textbf{Co-design for multiple OCT design parameters:} The proposed co-design framework (as shown in Fig. \ref{fig:OCT co-design}) evaluates multiple design parameters at the same time in relation to the maximum power-to-weight for the OCT system when operating in the spatial-temporal uncertain ocean environment. 
\end{itemize}

\subsection{Results and Discussions}
\begin{figure}[t]
    \begin{center}
    \includegraphics[width=0.95\linewidth]{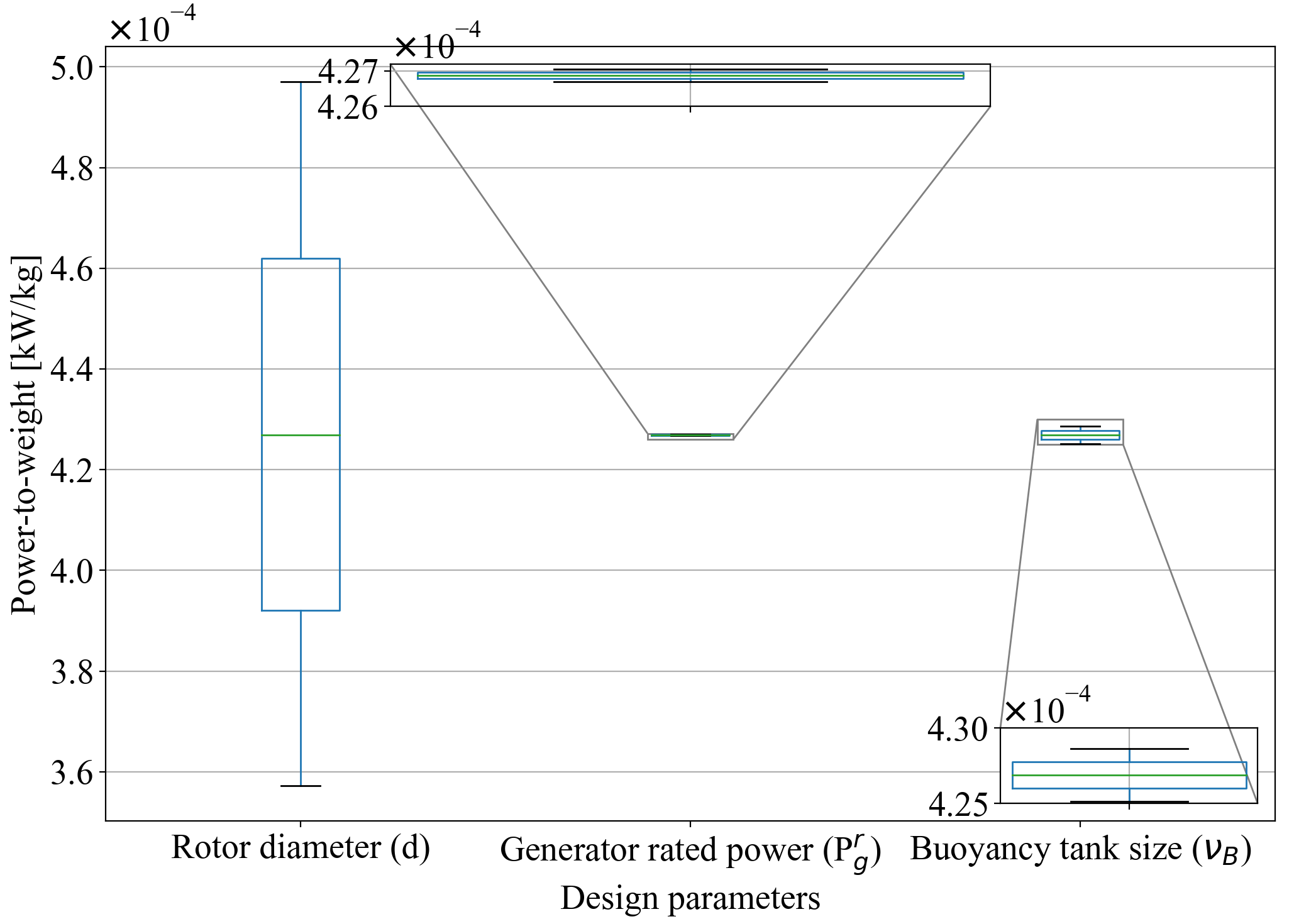}
    \end{center}
    \caption{Comparing the power-to-weight ratio changes due to changes ($\pm10$ \% from its base value) in design parameters, including rotor, generator, and buoyancy tank.}\label{fig:Sensitivity analysis}
\end{figure}

\begin{figure}[t]
    \begin{center}
    \includegraphics[width=0.95\linewidth]{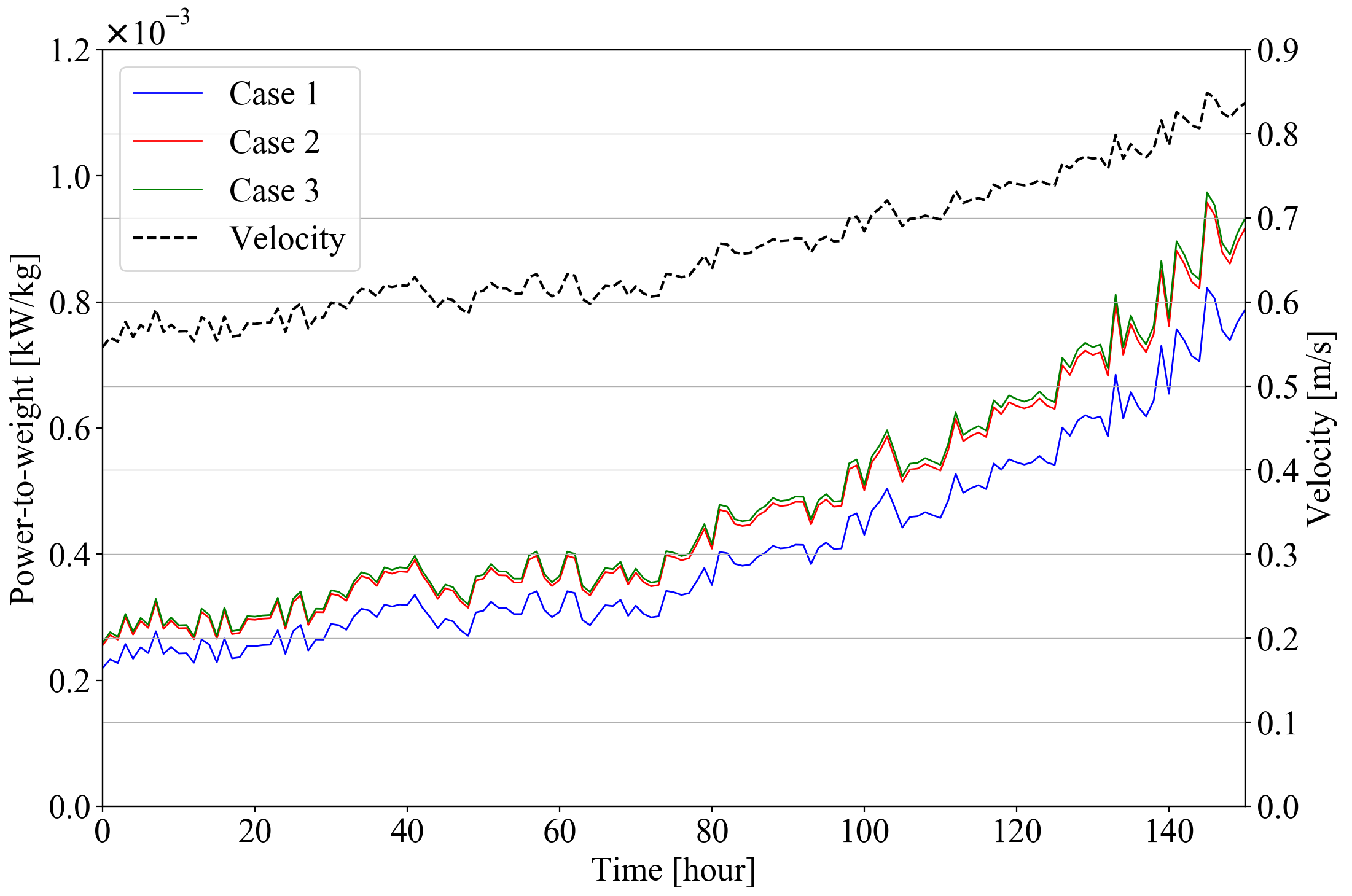}
    \end{center}
    \caption{Comparing optimal power-to-weight ratios obtained by (case I) baseline OCT design, (case II) co-design for the rotor, and (case III) co-design for multiple design parameters. The results are shown over a sample real measured ocean current velocity from the Gulf Stream.}\label{fig:optimal power-to-weight} 
\end{figure}

Fig. \ref{fig:Sensitivity analysis} compares the obtained power-to-weight ratios for changes applied in each design parameter ($\pm10$ \% from its base value). From this figure, the rotor diameter is the dominant design parameter, which significantly affects the power-to-weight ratio. The design parameters of the baseline OCT, the optimal design parameters obtained through co-design for the rotor, and co-design for multiple parameters simultaneously are presented in Table \ref{table:Robustness}. The hourly power-to-weight ratios for baseline design and optimal designs for the other two cases, along with the corresponding ocean current velocity (i.e., the turbine operating environment), are shown in Fig. \ref{fig:optimal power-to-weight}. The optimal OCT designs for case II and case III generate the same power (e.g., $P(\upsilon)=256.99$ $\mathrm{kW}$) but different total mass. The rotor diameter dominates other design parameters due to its significant mass and effect on the power generation ($P_{\mathrm{OCT}}$). From all these three cases, we see that the generated powers $P(\upsilon)$ are smaller than the baseline design of 700 $\mathrm{kW}$, which means that the baseline design was away from the optimal. The reason is that the average ocean current speed that we used in this paper is around 0.67 $\mathrm{m/s}$, much smaller than the speed considered for formulating the baseline design 1.015 $\mathrm{m/s}$. Through control co-design, the optimal generator rated power in case III is largely decreased from 700 $\mathrm{kW}$ in baseline design to 495 $\mathrm{kW}$, much closer to the harvesting power 256.99 $\mathrm{kW}$. These results highlight the importance of applying the geometric/spatial-temporal control co-design approach to design the optimal OCT in dynamic ocean environments.

\begin{table}[t]
    \caption{Comparing the optimal OCT designs obtained through three cases: (case I) baseline OCT design, (case II) co-design for the rotor, and (case III) co-design for multiple design parameters.}
\centering
    \begin{tabular}{|c|c|c|c|c|}
        \hline
        Parameter & Unit & Case I & Case II & Case III\\
         \hline
        $d_\mathrm{r}$ & $\mathrm{m}$ & 20 & 22 & 22\\
        $P^{\mathrm{r}}_{\mathrm{g}}$ & $\mathrm{kW}$ & 700 & 700 & 495\\        
        $\nu_{\mathrm{B}}$ & $\mathrm{m^{3}}$ & 31.215 & 31.215 & 18.824\\
        $m_{\mathrm{r}}$ & $\mathrm{kg}$ & 61191 & 80795 & 80795\\
        $m_{\mathrm{g}}$ & $\mathrm{kg}$ & 2247 & 2247 & 1633\\
        $m_{\mathrm{B}}$ & $\mathrm{kg}$ & 20427 & 20427 & 12372\\ 
        $m_{\mathrm{t}}$ & $\mathrm{kg}$ & 497418 & 517021 & 508353\\ 
        $P(\upsilon)$ & $\mathrm{kW}$ & 212.34 & 256.99 & 256.99\\
        $\frac{P(\upsilon)}{m_{\mathrm{t}}}$ & $\mathrm{kW/kg}$ & 4.269e-4 & 4.971e-4 & 5.055e-4\\
        \hline
    \end{tabular}
    \label{table:Robustness}
\end{table}

This work is the first attempt to control co-design for a buoyancy-controlled OCT system. The efficacy of the proposed approach, which maximizes the power-to-weight ratio and finds the optimal values of three key design parameters, was validated through simulations. The current study can be expanded on different aspects. The cost minimization objective should be added to the control co-design framework in addition to the optimization of the power-to-weight ratio. Furthermore, the spatial-temporal optimization in the current format ignores the complete linear model of the OCT and, in a broader view, a dynamic model of the OCT as well as the coupling between all OCT states. Although the presented model takes into account measured ocean currents, the increased forcing on different components is neglected, which is bounded by defining an upper limit for changing the design parameters. Finally, the ultimate goal of control co-design is to optimize the whole OCT design parameters with detailed modeling of the corresponding couplings.

\section{CONCLUSION} \label{sec:conclusion}
In this paper, a control co-design framework for a buoyancy-controlled OCT was presented to take into account the spatial-temporal path planning and turbine geometry when maximizing the power-to-weight ratio. The investigated framework accounts for key design parameters, including the sizes of the generator, rotor, and variable buoyancy tank. In the current study, we assumed an ideal case to maintain the baseline dimensions for the OCT system while changing the design parameters. Future work is required to fully investigate the relation and coupling between different design parameters and components; also, other critical design parameters (such as mooring cable) should be studied in the co-design framework.





\bibliography{main}
\bibliographystyle{ieeetr}


\end{document}